\documentclass{iau}

\usepackage{amsmath}
\usepackage{graphicx}
\usepackage{multirow}
\usepackage{wrapfig}

\begin{document}

\lefttitle{Cambridge Author}
\righttitle{Coronal Mass Ejections from Young Suns: Insights from Solar and Stellar Observations and Models}

\jnlPage{1}{7}
\jnlDoiYr{2021}
\doival{10.1017/xxxxx}

\aopheadtitle{Proceedings IAU Symposium}
\editors{eds.}

\title{Coronal Mass Ejections from Young Suns: Insights from Solar and Stellar Observations and Models}

\author{Kosuke Namekata}
\affiliation{Heliophysics Science Division, NASA Goddard Space Flight Center, 8800 Greenbelt Road, Greenbelt, MD 20771, USA}
\affiliation{The Catholic University of America, 620 Michigan Avenue, N.E. Washington, DC 20064, USA}
\affiliation{The Hakubi Center for Advanced Research, Kyoto University, Yoshida-Honmachi, Sakyo-ku, Kyoto 606-8501, Japan}
\affiliation{Department of Physics, Kyoto University, Kitashirakawa-Oiwake-cho, Sakyo-ku, Kyoto, 606-8502, Japan}

\begin{abstract}
Recent discoveries have revealed exoplanets orbiting young Sun-like stars, offering a window into the early solar system. These young stars frequently produce extreme magnetic explosions known as superflares potentially leading to fast and massive coronal mass ejections (CMEs). Recent research have highlighted the importance of stellar CMEs, as these events and associated particles can trigger atmospheric loss and initiate chemical reactions in planetary atmospheres. However, the observation of these associated CMEs remains largely unexplored, marking a crucial first step in assessing the particle environment. Here we present the results of 5-years multi-wavelength observations of young Sun-like stars, providing the critical clues to the common picture of solar and stellar CMEs. This comprehensive study suggests that further advancing the use of solar model could provide the first empirical inputs into calculations of atmospheric escape/chemical reactions for young planets.
\end{abstract}


\maketitle

\section{Introduction}

Solar flares are often accompanied by coronal mass ejections (CMEs), which are generally understood within the framework of magnetic reconnection models \citep{2011LRSP....8....6S}. 
Large CMEs have profound impacts on planetary environments and even pose risks to human civilization \citep{2022LRSP...19....2C}. 
The occurrence of CMEs is well known for the Sun, but remains poorly understood for other stars (see review by \citealt{2022arXiv221105506N,2025LRSP...22....2V}). The recent discovery of more than 6,000 exoplanets has underscored the importance of investigating stellar space weather, including stellar CMEs \citep{2020IJAsB..19..136A}. 
While many efforts to detect stellar CMEs have been directed toward M dwarfs (e.g., \citealt{2016A&A...590A..11V,2018PASJ...70...62H,2020PASJ..tmp..253M,Notsu2023,2024PASJ...76..175I,2025ApJ...979...93K}), young Sun-like stars (G dwarfs) are of particular importance as they provide natural analogues of the early Sun–Earth system, offering key insights into how Earth-like planetary atmospheres developed and how conditions for the emergence of life may have arisen. 
These young Sun-like stars exhibit extremely high levels of magnetic activity, frequently producing so-called ``superflares" with energies exceeding 10$^{33}$ erg \citep[e.g.,][]{2007LRSP....4....3G,2012Natur.485..478M,2019ApJ...876...58N,2020arXiv201102117O,2022ApJ...926L...5N}. A key open question is whether such energetic events are accompanied by massive CMEs capable of substantially influencing planetary atmospheres and even stellar mass and angular momentum evolution. 

The primary objective of our project is to characterize stellar CMEs from young Sun-like stars. To this end, we have been conducting a five-year (132-night) multi-wavelength monitoring campaign since 2020, taking advantage of the flexible scheduling capabilities of the 3.8m Seimei Telescope/KOOLS-IFU (R$\sim$2000) in Japan. Our strategy is to identify Doppler-shifted signatures in the H$\alpha$ line as indicators of filament and prominence eruptions.
We focus on two nearby G dwarfs, EK Dra (G1.5V, age $\sim$50-125 Myr) and V889 Her (G0V, age $\sim$25 Myr), which are among the best proxies for the young Sun. 
These data are complemented by optical photometry from TESS (600-1000 nm), soft X-ray monitoring with NICER (0.5–3 keV), and magnetic field mapping through Zeeman Doppler imaging with TBL/NARVAL in France. Together, these coordinated observations provide an unprecedented dataset for investigating the occurrence and properties of CMEs on young Sun-like stars.

\section{Discovery of Stellar Filament Eruptions}

\begin{figure}[!htb]
  \vspace*{-0.0 cm}
\begin{center}
\includegraphics[width=5.in]{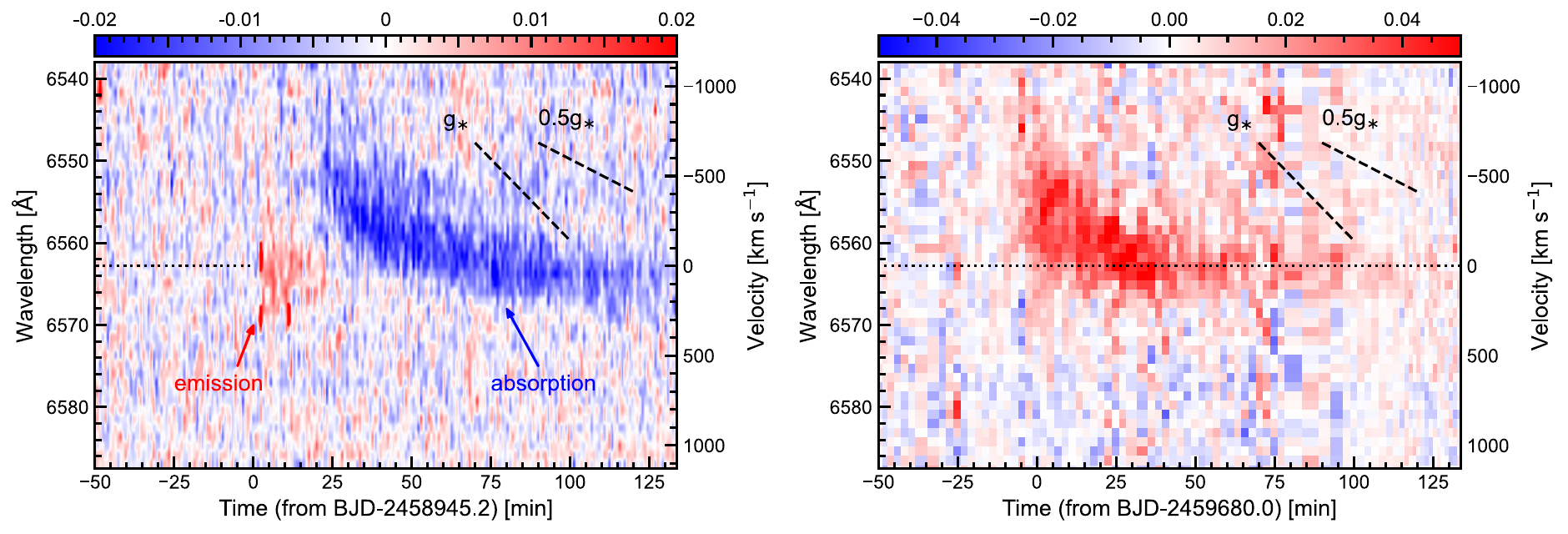} 
\caption{Dynamic spectrum of H$\alpha$ line. Left: filament eruption (blueshifted absorption, adapted from \citealt{2022NatAs...6..241N}); Right: prominence eruption (blueshifted emission, \citealt{2024ApJ...961...23N})}
\label{fig:1}
\vspace*{-1.0 cm}
\end{center}
\end{figure}

First we succeeded in detecting a superflare on EK Dra with an energy of $2 \times 10^{33}$~erg through simultaneous H$\alpha$ spectroscopy and TESS white-light photometry on April 5, 2020. 
In the post-flare phase, we detected a blueshifted absorption component with a velocity of $510~\mathrm{km~s^{-1}}$ (Figure \ref{fig:1}, left), providing the first clear evidence of a stellar filament eruption on a Sun-like star \citep{2022NatAs...6..241N}. To address the question of what such stellar eruptions look like, we performed a Sun-as-a-star analysis of H$\alpha$ line profiles during solar eruptive events. The spectral variations observed in these solar eruptions closely resemble those identified in EK~Dra, leading us to conclude that stellar filament eruptions represent large-scale analogues of solar filament eruptions.
A pseudo two-dimensional (2D) MHD simulation by \cite{2024ApJ...963...50I} also supports this view and suggests that it can eventually lead to steller CMEs.
Compared to typical solar eruptions (a few hundred km s$^{-1}$ and $\lesssim10^{17}$ g), stellar eruptions are faster, more massive, and carry greater kinetic energy, indicating that their impact on interplanetary space would be far more severe than what humans have experienced on the present-day Earth.

\section{Multiwavelength Signatures of Stellar CMEs}\label{sec:3}

Another EK Dra's superflare on April 10, 2022 with an energy of $1.5 \times 10^{33}$~erg exhibited a blueshifted H$\alpha$ emission component with velocities of $400$--$690~\mathrm{km~s^{-1}}$ (Figure \ref{fig:1}, right), indicating an off-limb prominence eruption extending beyond the stellar surface \citep{2024ApJ...961...23N}. Nearly simultaneous NICER X-ray observations revealed possible coronal X-ray dimming, which is a different CME signature representing the evacuation of coronal material \citep{2016SoPh..291.1761H,2021NatAs...5..697V}, detected about two hours later of the H$\alpha$ prominence eruption \citep{2024ApJ...961...23N}. 
Although the possible coronal dimming is just a upper limit of evacuated mass, it is intriguing to see that the estimated prominence mass ($\sim 10^{19}$--$10^{20}$~g) was significantly larger than the evacuated coronal mass ($\sim 10^{17}$--$10^{18}$~g).

\section{Data-Driven Models}\label{sec:4}

\begin{figure}[!htb]
  \vspace*{-0.2 cm}
\begin{center}
\includegraphics[width=4.in]{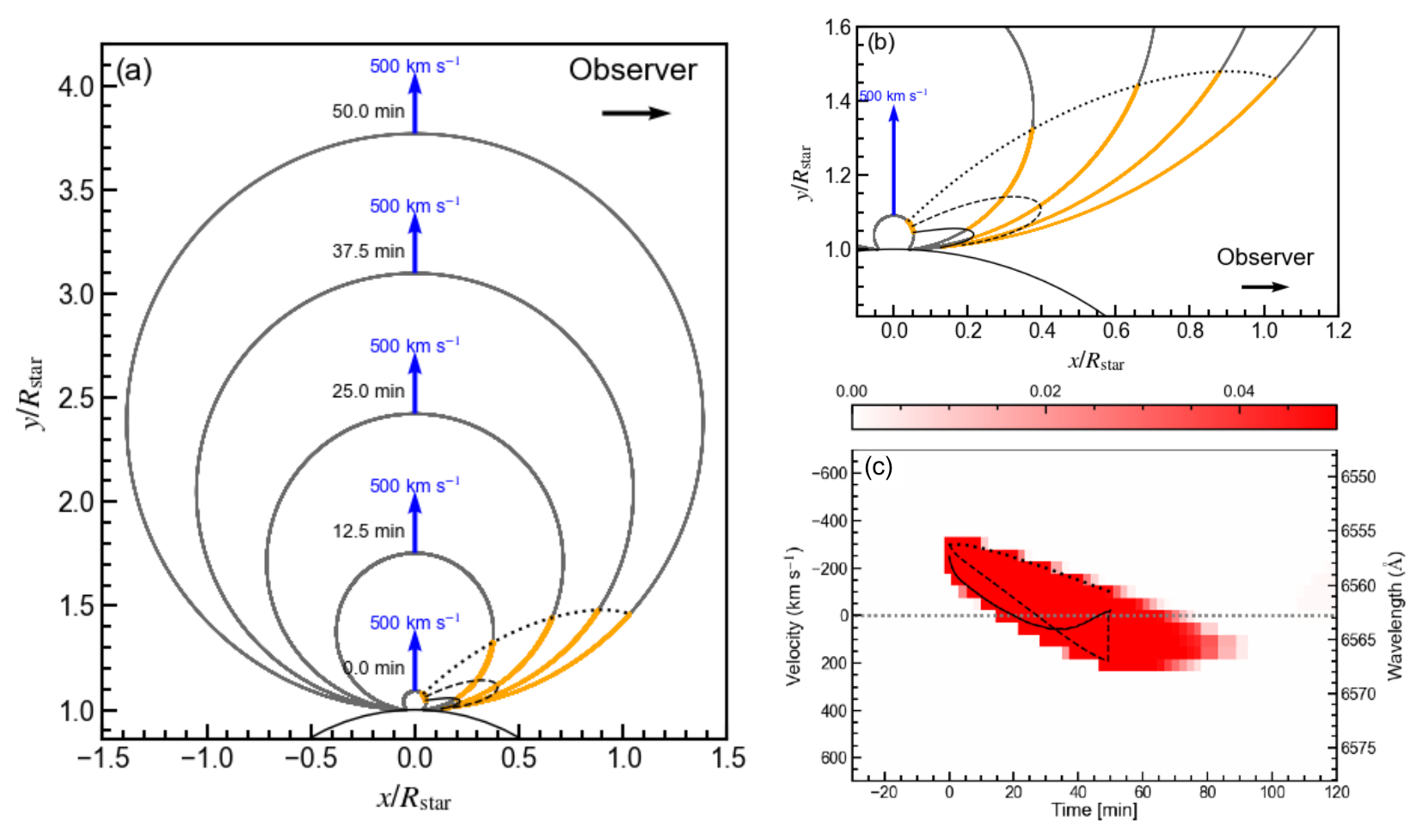} 
  \vspace*{-0.2 cm}
\caption{(a,b) Assumed geometry in a pseudo 2D MHD model of magnetic loop expantion (black lines) and prominence material (orange). (c) Simulated H$\alpha$ dynamic spectrum.}
\label{fig:2}
\vspace*{-1.0 cm}
\end{center}
\end{figure}

Two data-driven models were employed to investigate the prominence eruption in Section \ref{sec:3} \citep{2024ApJ...976..255N}. One method is the one-dimensional free-fall model. A parameter survey of ejection angles using this simple model indicated that the prominence was launched at an angle of approximately $20^{\circ}$ from near the stellar limb. 
Building on this result, we performed a pseudo 2D MHD simulation based on \cite{2024ApJ...963...50I}'s model, employing a self-similar expansion model based on solar physics to describe the erupting prominence, as in Figure \ref{fig:2}(a,b). 
The simulation successfully reproduced the observed H$\alpha$ spectra (Figure \ref{fig:2}(c)), suggesting that stellar prominence eruptions can be explained by solar model, and while the prominence itself decelerates, the magnetic loop can continue to expand to be stellar CMEs. 

\section{Magnetic Environment of Eruptions}

\begin{wrapfigure}{r}{0.45\textwidth}
\vspace{-1.2cm}
\centering 
\includegraphics[width=0.45\textwidth]{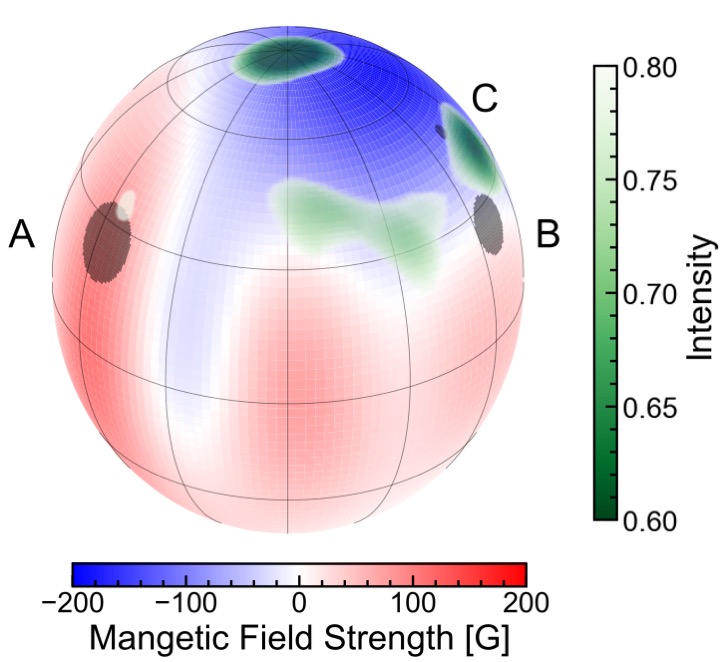}
\vspace{-0.6cm}
\caption{Starspot maps when the prominence eruption in Figure \ref{fig:1} (right) occurred, showing the Doppler imaging (green) and those from the TESS light-curve modeling (gray circles), the radial magnetic field (red and blue colors).}
\label{fig:3}
\vspace{-0.3cm}
\end{wrapfigure}

As for the event discussed in Sections~\ref{sec:3} and \ref{sec:4}, we performed a simultaneous analysis combining \textit{TESS} light-curve inversion and Zeeman Doppler Imaging \citep{2024ApJ...976..255N}. 
The resulting Figure \ref{fig:3} displays the large-scale magnetic field configuration together with the starspot distribution during the period of the prominence eruption. 
This analysis reveals the consistent presence of starspots, such as Spot~A, Spot~B, and Spot~C, located at mid-latitudes. Notably, Spot~B lies close to the polarity inversion line. If the prominence eruption originated from starspots, these mid-latitude spots are plausible source regions. However, this region is dominated by closed magnetic fields, raising another possibility that the eruption originated from the open-field regions near the pole.


\section{Frequency and CME-Driven Mass Loss}

As a result of five-year dedicated observations, we obtained 15 H$\alpha$ superflares. Among those, we found two filament eruptions and two prominence eruptions \citep{2025.namekata.apj}. The lower limit of the eruption-flare association rate is 27$_{-16}^{+25}$\%, leading occurrence rates of 0.21$_{\pm0.12}$ and $<$0.32$^{+0.46}_{-0.32}$ events/day. We estimated the lower limit of the CME-driven mass-loss rate for EK Dra as $4 \times (10^{-13}$--$10^{-12})$ $M_{\odot}$ yr$^{-1}$.
This is comparable to the stellar wind mass loss at a similar age, suggesting that we would likely need to take CME mass loss into account in the context of stellar mass/angular momentum evolution.

\section{Summary and Conclusion}

Through five years of dedicated observations in the framework of international collaborations, we have obtained the first evidence of stellar filament/prominence eruptions from a young Sun-like star. These efforts have also enabled us to derive their statistical properties for the first time, and to identify multi-wavelength signatures of stellar CME activity. We conclude that superflares on young Sun-like stars are likely associated with fast, massive, and frequent solar-like eruptions. These findings provide new insights into the environment of the early Sun at an age of $\sim$100 Myr, and suggest that such activity may have profoundly influenced the atmospheres and habitability of the early Earth, Mars, and Venus.

\noindent{\bf Acknowledgment} The author acknowledges all collaborators who contributed to these works.

\bibliography{Sample}{}
\bibliographystyle{aasjournal}

\end{document}